\documentstyle[12pt]{article}
\begin{document}
\thispagestyle{empty}
\newcommand{\om}{{\stackrel{(1)}{\omega}}_n}
\newcommand{\omn}{{\stackrel{(1)}{\omega}}_{-n}}
\newcommand{\omm}{{\stackrel{(2)}{\omega}}_n}
\newcommand{\ommn}{{\stackrel{(2)}{\omega}}_{-n}}
\newcommand{\bb}{\mbox {\boldmath $\beta $}}
\newcommand{\ab}{\mbox {\boldmath $\alpha $}}
\renewcommand{\theequation}{\arabic{section}.\arabic{equation}}
\begin{flushright}
{\large \bf JINR preprint E2-93-193\\
Dubna, 1993}
\end{flushright}
\vskip0.8cm
\begin{center}
{\large \bf  Is  it  possible  to  assign  physical meaning to field
theory with higher derivatives?} \\[0.8cm]
{\bf A.M.  Chervyakov$\dagger  $ and V.V.  Nesterenko$\ddagger $} \\
$\dagger $Laboratory of Computing Techniques  \&  Automation,  Joint
Institute for Nuclear Research, Dubna SU-141980, RUSSIA \\
E-mail address: chervyakov@main1.jinr.dubna.su
$\ddagger $Laboratory  of  Theoretical Physics,  Joint Institute for
Nuclear Research,\\ Dubna SU-141980, RUSSIA \\
E-mail address: nestr@theor.jinrc.dubna.su

\vspace*{1cm}
            {\large \bf Abstract }\\[0.5cm]
\end{center}
To overcome the difficulties with the energy indefiniteness in field
theories  with  higher  derivatives,  it  is  supposed  to  use  the
mechanical analogy, the Timoshenko theory of the transverse flexural
vibrations of beams or rods well known in mechanical engineering. It
enables one to introduce the notion of a "mechanical" energy in such
field models that is wittingly positive definite.  This approach can
be  applied  at  least  to  the  higher  derivative   models   which
effectively describe the extended localized solutions in usual first
order field theories (vortex solutions in Higgs models and  so  on).
Any  problems  with  a  negative  norm  ghost  states  and unitarity
violation do not arise here.

\newpage
\vspace*{2cm}
\section{Introduction}
Field theories  with  higher derivatives acquire a stable reputation
of  nonphysical  theories.  Nevertheless,  because  of  they   being
frequently  arise  in  different  areas  of  theoretical physics the
interest in this issue is periodically revived [1--10].

  A principal  shortcoming  of  higher  derivative  theories,   both
classical and quantum,  is the lack of lower--energy bound. Here the
energy is implied as a conserved Noether quantity  corresponding  to
the  translation  invariance  of the theory with respect to time or,
that is  the  same,  as  a  value  of  the  Hamiltonian  constructed
according  to  Ostrogradsky's rules on the solution of the equations
of motion [11].

The attractive properties of the quantum field theories with  higher
derivatives is also worth mentioning. In particular, the convergence
of Feynman diagrams is improved owing to the higher derivative terms
in  Lagrangian.  For  example,  the conformal gravity is found to be
renormalizable whereas the Einstein one is not [4,  12].  Just  this
property  of  theories  in  question  is used to construct the gauge
invariant renormalization of Yang--Mills fields by adding the higher
derivative terms to the standard Lagrangian [13].

It should  be  noted  that  the  lack  of  lower  energy bound for a
completely isolated system is admissible in principle if the  energy
is  an  integral of motion.\footnote{It is usually believed that the
energy being indefinite in  sign  entails  the  instability  of  the
classical  dynamics  for theories with higher derivatives,  although
the very special counterexample is known [14].  More exactly, if the
energy of a system is not definite in sign, the problem of stability
cannot be solved using the Lagrange--Dirichlet theorem [15] and,  in
general, it is not reduced to searching for the Lyapunov function as
in the  case  of  the  usual  theories  with  Lagrangian  functions,
containing,  at  most,  the  first  derivative  in time of dynamical
variables.}  But,  unfortunately,  such  isolated  systems  are  not
realized practically.  Nonremovable  interaction  with  an  external
environment inevitably results in pumping out an arbitrary amount of
the energy from the system, lowering its energy without limits.

Obviously, the  higher derivatives in time in the Lagrangian lead to
additional  degrees  of  freedom,  since   there   is   one--to--one
correspondence  between  the  dynamical  degrees  of freedom and the
initial data for the  relevant  Euler--Lagrange  equations.  In  the
following,  for  the sake of definiteness we shall discuss the field
theories with Lagrangian functions depending, at most, on the second
derivatives in time.  Here there arises the very typical picture for
higher derivative theories:  besides the basic mode of  oscillations
which  takes  place even in the absence of the second derivatives in
Lagrangian there emerges additional,  as  a  rule,  higher-frequency
mode.  The  contribution  to  the  energy of the second mode has the
opposite sign as compared with the basic one. Therefore, even at the
classical  level it turns out to be more profitable energetically to
excite the oscillations from the second mode.  The more oscillations
of  that  sort are excited and the larger their amplitudes are,  the
lower the total energy of a system turns out to  be.  From  this  it
follows   that  the  field  theories  with  higher  derivatives  are
unacceptable physically at least in making  use  of  their  standard
interpretation.

All these  arguments  are  applied  exactly  to the quantum level as
well.  Here the oscillations of both positive  and  negative--energy
modes  are  associated with the corresponding quanta of excitations.
In  virtue  of  the   impossibility   of   removing   the   external
perturbations,  as it has been noted previously, an unlimited number
of the negative energy quanta will be created.  As a result,  in the
field  theories with higher derivatives a problem alike the infrared
catastrophe  in  quantum  electrodynamics  arises,   but   for   all
frequencies  of  the second mode now.  This problem was successfully
overcome in electrodynamics,  but it still remains unsolved  in  the
higher derivative theories.

Some time  ago,  it  was  popular  to  use  here  the  formalism  of
indefinite metric in the Fock space of the states.  This metric  can
be introduced by mutual interchange of the creation and annihilation
operators of quanta of the second mode.  As a  result,  the  quantum
states with excitations from the second mode acquire a negative norm
but  the  energy  calculated  as  an  expectation   value   of   the
Ostrogradsky  Hamiltonian  over  these  states  turns  out  to  be a
positive definite quantity [1,  2].  Thereby,  the  problem  of  the
negative   energy   is   reduced   to  searching  for  the  physical
interpretation of theories with implicit-violated unitarity.  So far
there  is  no acceptable solution of the problem along this way [9].
Therefore in the following we shall only deal with the difficulty of
the  energy  being  indefinite  in  sign in the theories with higher
derivatives.

As far as we know,  the attempts to attach the physical  meaning  to
the   higher  derivatives  theories  are  based  on  the  conjecture
forbidding the excitations with  negative  energy.  This  constraint
should appear as the boundary condition following from the cosmology
[7] or as a by-product of the nonperturbative quantum solutions [5],
or  it  has  been  introduced  from  the outset in formulating these
models [10].

We would like to suggest another solution of the problem. Namely, we
will  show that the energy in the theory with higher derivatives can
be redefined using a mechanical analogy.  Here we have in  mind  the
special  class  of  higher  derivative  theories  arising  when  the
effective Lagrangians are  constructed  in  extended  object  models
(strings,  in  particular).  Even at the classical level an extended
object requires the field description.  We shall  suppose  that  the
original  field  theory does not contain the higher derivative terms
in the Lagrangian so that its energy  is  bounded  from  below.  The
neglect  of the details of internal structure of the extended object
along one or several its internal dimensions results,  as a rule, in
higher derivative terms in the effective Lagrangian.  Now the energy
of the effective theory turns out to be unbounded from below.

As a specific model, we shall treat a relativistic rigid string with
the  action functional depending on the second derivatives of string
coordinates [16,  17]. Here the rigidity term takes effectively into
account  the  thickness  of the string.  It may be imagined clearly,
that this system simulates,  for example,  the gluon tube of  finite
radius  that  connects  the  quarks into the hadrons.  Such a simple
picture arises in certain approximations to QCD [18, 19]. Taking the
finite  thickness  of the cosmic strings into account one arrives at
the model of the rigid string as well [20--22].

To solve the equations of motion in the model  of  the  relativistic
string   with   rigidity,  we  confine  ourselves  to  the  harmonic
approximation in the timelike  gauge.\footnote{  As  is  known,  the
total  theory  of the relativistic string with rigidity owing to the
reparametrization invariance of its action  is  a  dynamical  system
with  constraints  in the phase space [23].  However,  the number of
these constraints is not enough to remove all the quanta of negative
energy.} Then we shall elucidate an analogy between the rigid string
and the most simple mechanical system that takes  into  account  the
stiffness  of  an  extended  vibrating  body.  To  this end we shall
consider Timoshenko's theory of the flexural vibrations of beams and
rods  well  known  in  the mechanical engineering [24].  This theory
takes effectively into account the finite thickness of the beam  via
the second derivatives in time and in longitudinal coordinate in the
Lagrange function. It is important that there is no problem with the
energy being positive definite in this mechanical system. Thus, this
analogy points out in what way the definition of the energy  in  the
model  of  the  rigid  string  should  be  changed to get a positive
definite energy.

The outline of the paper is as follows.  In Sec.\ II the problem  of
the  energy  unbounded from below typical of the field theories with
higher derivatives is discussed in the framework of the relativistic
string  with  rigidity  by making use of the harmonic approximation.
The theory of flexural vibrations of the Timoshenko beam is given in
Sec.\   III.  The  Hamiltonian  constructed  here  by  applying  the
Ostrogradsky  rules  leads  to  the  energy  unbounded  from  below.
Nevertheless  in this case there exists the notion of the mechanical
energy which is positive definite.  In Sec.\ IV the analogy  between
the  relativistic  string  with rigidity and the beam or rod is used
for constructing the  positive  definite  energy.  In  Sec.\  V  the
proposed  method  is  compared  with  other attempts to overcome the
drawback related with the energy unbounded from below in  the  field
theories with higher derivatives.

\section{ Harmonic appriximation in the rigid string model}
\setcounter{equation}0
  The localized  vortex  solutions  to  the  classical  equations of
motion having the form of a flux tube or a string are well known  in
the  gauge  fields  models  with the Higgs Lagrangian [25--27].  The
behaviour of these solutions can  be  described  by  some  effective
Lagrangians [22,  28]. In the zeroth order approximation in the flux
tube  width  one  obtains  here  the  Nambu--Goto  action  for   the
relativistic  string  [18].  The  first order correction in the tube
width leads to a rigid string model with the action depending on the
second derivatives of the string coordinates [20, 21]
\begin{equation}
W\,=\,-\rho_{0}c\int\int d^{2}u\sqrt{-g}\left(1\,-\,\frac{\alpha}
{2}r^2_s\Delta x^\mu\Delta x_\mu\right).
\end{equation}
Here $x^{\mu}(u^0,u^1)\,,\,\mu=0,1,\ldots,D-1,$   are   the   string
coordinates in the $D$--dimensional space--time whose metric has the
signature $(+,-,\ldots,-)\,,\, \rho_0$ is the linear mass density of
the  flux  tube (or of the string),  $r_s$ is the transverse size of
this tube,  $c$ is the velocity of light.  The internal geometry  on
the   string   world  surface  is  defined  by  the  induced  metric
$g_{ij}(u)\,=\,\partial_{i}x^{\mu}\partial_{j}x_{\mu}\,,\,
i,j=0,1\,,\,g\,=\,\det(g_{ij})\,,\,g\,<\,0$.  The  Laplace--Beltrami
operator with respect to this metric reads explicitly
\begin{equation}
\Delta\,=\,\frac{1}{\sqrt{-g}}\frac{\partial}{\partial
u^i}\left(\sqrt{-g}    g^{ij}\frac{\partial}{\partial   u^j}\right),
\quad g_{ij}g^{jk}\,=\,\delta_i^k.
\end{equation}
For the curvilinear coordinates $u^0$ and $u^1$ on the string  world
sheet  we  shall  frequently  use another,  more ordinary,  notation
$u^0\,=\,\tau\,,\,u^1\,=\, \sigma$. The numerical parameter $\alpha$
in   the  action  (2.1)  is  specified  by  the  concrete  mechanism
generating the flux tube.  In  the  abelian  gauge  model  with  the
simplest  Higgs  potential (the Nielsen--Olesen vortex model for the
relativistic string) $\alpha$ proves to be about 20 [20]. The action
(2.1)  results  in  the nonlinear equations of motion containing the
partial derivatives of the fourth order of  the  string  coordinates
$x^\mu$  [29].  To  advance in their study,  we employ the following
parametrization including the time--like gauge on the  string  world
surface

\begin{equation}
x^{\mu}(u)\,=\,\left\{ct,\;\frac{l}{\pi}\sigma,\;{\bf x}(u)\right\},
\quad \tau \,=\,t,
\end{equation}
where ${\bf   x}(u)$  are  $(D-2)$  transverse  string  coordinates.
Although the parametrization (2.3) holds true only for  the  limited
string motions (so--called harmonic approximation [30]),  it will be
sufficient for our aims.

Inserting the ansatz (2.3) into (2.1) and expanding the integrand of
(2.1)  up  to second order terms in powers of ${\bf x}(u)$ we obtain
[30]
\begin{equation}
W\,=\,\frac{\rho_{0}}{2\pi}\int dt\int\limits_{0}^{\pi}d\sigma
\left[\dot{
{\bf x}}^2\,-\,a^{2}{\bf x}'^2\,-\,\epsilon a^2\left(a^{-2}
\ddot{{\bf x}}\,-\,
{\bf x}''\right)^2\right],
\end{equation}
where $a\,=\,\pi                  c/l,\,\epsilon\,=\,\alpha\left(\pi
r_{s}/l\right)^2$,  $l$  is  the  string  length.  The   dot   means
differentiation  with  respect  to $t\,=,\tau $ and the prime,  with
respect to $\sigma $.  Variation  of  the  action  (2.4)  gives  the
following equations of motion
\begin{equation}
\left(1\,+\, \epsilon\Box\right)\Box{\bf x}(u)\,=\,0,
\end{equation}
\[\Box\,\equiv\,a^{-2}\frac{\partial^2}{\partial t^2}\,-\,
\frac{\partial^2}
{\partial \sigma^2}\]
and the boundary conditions
\begin{equation}
\left(1\,+\,\epsilon\Box\right){\bf x}'\,=\,0,
\end{equation}
\[\Box{\bf x}\,=\,0,\quad
\sigma\,=\,0,\pi.\]
Owing to equations  (2.5)  and  (2.6)  being  linear  their  general
solution can be represented as the sum
\begin{equation}
{\bf x}(t,\sigma)\,=\, {\bf x}_{1}(t,\sigma)\,+\,{\bf x}_{2}
(t,\sigma).
\end{equation}
Here ${\bf x}_1(u)$ are transverse degrees of freedom of the open
Nambu--Goto string [18]
\begin{equation}
\Box{\bf x}_1(u)\,=\,0,
\end{equation}
\[{\bf x}'_1\,=\,0,\quad\sigma\,=\,0,\pi.\]
The coordinates ${\bf x}_{2}(u)$ obey the following equations
\begin{equation}
\left(1\,+\,\epsilon\Box\right){\bf x}_2(u)\,=\,0,
\end{equation}
\[{\bf x}_2\,=\,0,\quad\sigma\,=\,0,\pi.\]
As usual,  the  general  solution of the boundary problems (2.8) and
(2.9) is given by the  expansions  in  corresponding  eigenfunctions
[30]
\begin{equation}
{\bf x}_1(t,\sigma)\,=\,{\bf Q}\,+\,\frac{{\bf P}t}{\rho_{0}l}\,+
\,i\sqrt
{\frac{\hbar}{\pi\rho_{0}c}}\sum_{n\neq 0}\frac{\mbox
{\boldmath $\alpha $}_n}
{\om }\,\cos n\sigma \,e^{-ia\om t},
\end{equation}
\[
{\bf x}_2(t,\sigma)\,=\,\sqrt{\frac{\hbar}{\pi\rho_{0}c}}
\sum_{n\neq 0}
\frac{\mbox {\boldmath $\beta $}_n}{\omm }\,\sin n
\sigma \,e^{ia\omm t},
\]
with two series of the  eigenfrequencies
\begin{equation}
\om \,=\,-\omn =\,n,\quad
\omm \,=\,-\ommn \,=\,\sqrt{n^2\,+\,\frac{1}{\epsilon}},
\end{equation}
\[n\,=\,1,2,\ldots,.\]
Here ${\bf  Q}$  and  ${\bf P}$ are the coordinates of the center of
mass and the total momentum of the  string,  respectively,  and  the
amplitudes  ${\bf\alpha}_n$  and ${\bf\beta}_n$ in virtue of reality
of the variables ${\bf x}_1$ and ${\bf x}_2$ obey the usual rules of
complex conjugation
\begin{equation}
\ab ^{*}_{n}\,=\,\ab _{-n},\quad \bb ^{*}_{n}\,=\,
\bb _{-n},\quad n\,=\,0,\pm 1,\pm 2,\ldots.
\end{equation}

Thus, the transverse coordinates of  the  relativistic  string  with
rigidity ${\bf x}(u)$ are described in the harmonic approximation by
the   pair   of   independent   variables   $\left({\bf    x}_1,{\bf
x}_2\right)$. This duplication of the number of dynamical degrees of
freedom is general  for  higher  derivative  theories.  It  is  also
reflected  explicitly  in  the  canonical  formalism  worked out for
higher derivative theories  by  Ostrogradsky  more  than  a  centure
ago~[11]. In our case according to the Ostrogradsky method
the independent generalized coordinates are ${\bf q}_1\,=\,{\bf x}$
and ${\bf q}_2\,=\,\dot{{\bf x}}$ and their  conjugate  momenta  are
defined by the expressions
$$
{\bf p}_1\,=\,\frac{\partial {\cal L}}{\partial\dot{{\bf x}}}\,-\,
\frac{\partial}
{\partial t}\left(\frac{\partial {\cal L}}{\partial \ddot{{\bf x}}}
\right)\,=\,
\frac{\rho_{0}l}{\pi}\left(1\,+\,\epsilon\Box\right)\dot{{\bf x}},
$$
\begin{equation}
{\bf p}_2\,=\, \frac{\partial {\cal L}}{\partial \ddot{{\bf x}}}\,=
\,-\epsilon\frac{\rho_{0}l}{\pi}\Box{\bf x}.
\end{equation}
With the  use  of  (2.7),  (2.8) and (2.9) from (2.13) we find ${\bf
p}_1\,=\,  \left(\rho_{0}l/\pi\right)\dot{{\bf  x}}_{1},\quad   {\bf
p}_2\,=\,\left(  \rho_{0}l/\pi\right){\bf  x}_2$.  As a result,  the
canonical Ostrogradsky Hamiltonian
\begin{equation}
H\,=\,\frac{\rho_{0}l}{2\pi}\int\limits_{0}^{\pi}d\sigma
\left({\bf p}_1
\dot{{\bf x}}\,+\,{\bf p}_2\ddot{{\bf x}}\,-\,{\cal L}\right)
\end{equation}
in terms of the variables ${\bf x}_1$ and ${\bf x}_2$ takes the form
\begin{equation}
H\,=\,\frac{\rho_{0}l}{2\pi}\int\limits_{0}^{\pi}d\sigma\left[\left(
\dot{{\bf x}}^2_1\,+\,a^2{\bf x}'^2_1\right)\,-\,\left(
\dot{{\bf x}}^2_2\,-\,
a^2{\bf x}'^2_2\,-\,\frac{a^2}{\epsilon}{\bf x}^2_2\,-\,2{\bf x}_2
\ddot{{\bf x}}
_2\right)\right].
\end{equation}
 Hence it  follows  that  already  at  the   classical   level   the
excitations  of  the  degrees  of  freedom  ${\bf  x}_2$  may give a
negative contribution to the total energy  of  the  string.  Indeed,
inserting the general solution (2.10) into (2.15) we obtain
\begin{equation}
E\,=\,\frac{{\bf P}^2}{2M}\,+\,\frac{a\hbar}{2}\sum_{n=1}^{\infty}
\left(
{\ab }^{*}_{n}{\ab }_n\,+\,{\ab }_n{\ab }^{*}_{n}\right)\,
-\, \frac{a\hbar}{2}\sum_{n=1}^{\infty}\left({\bb }^{*}_{n}{\bb }_n\,
+\,{\bb }_n{\bb }^{*}_{n}\right),
\end{equation}
where $M\,=\,\rho_{0}l$ is the total mass of the string.

Thus, in the rigid string model we arrive at the problem general for
all  higher  derivative  theories  of  the  lack  of  lower   energy
bound~[10,  14].  In the quantum theory of this system the following
annihilation and creation operators $a^i_n$ and $b^i_n$ are defined
\[
\alpha^i_n\,=\,\sqrt{\om }\,a^i_n,\quad\alpha^{i}_{-n}\,=\,\alpha^
{+i}_{n}\,=\,\sqrt{\om }\,a^{+i}_{n},
\]
\[
\beta^i_n\,=\,\sqrt{\omm }b^i_n,\quad\beta^{i}_{-n}\,=\,\beta^{+i}_
{n}\,=\,\sqrt{\omm }b^{+i}_{n}
\]
with standard commutation relations
$$
\left[a^i_n,\,a^{+j}_{m}\right]\,=\,\left[b^i_n,\,b^{+j}_{m}\right]
\,=\,\delta
^{ij}\delta_{nm},
$$
$$
 i,j\,=\,1,2,\ldots,(D-2),\quad n,m\,=\,1,2,\ldots.
$$
Therefore, taking  account  of  the  zero--point oscillations of the
string we obtain the expression of the energy indefinite in sign
\begin{eqnarray}
E&=&\frac{{\bf P}^2}{2M}\,+\, a\hbar\sum_{n=1}^{\infty}\om \,
\left({\bf a}^{+}_{n}{\bf a}_{n}\,+\,\frac{D-2}{2}\right)\,-
\nonumber \\
&&-a\hbar\sum_{n=1}^{\infty}\omm \,\left({\bf b}^{+}_{n}
{\bf b}_{n}\,+\,
\frac{D-2}{2}\right).
\end{eqnarray}
As is well known [1,  2,  9], the negative energy $\left(-a\hbar\omm
\right)$ creation operators ${\bf b}^{+}_{n}$  can  be  regarded  as
positive   energy  $\left(+a\hbar\omm  \right)$  annihilation  ones.
Thereby,  in the Fock space of the states the positive norm negative
energy  excitations  are  transformed  into  negative  norm positive
energy ones. So, the violation of unitarity in the quantum theory is
really  reflection  of the essentially classical problem of the lack
of lower energy bound (see (2.16) and papers [9, 31, 32]).
In a  recent  papers  (see [10] for review) it was proposed to apply
the perturbative constraints to freeze out the excitations of  those
degrees of freedom which give rise to the negative contribution into
the energy.  In the persent paper using the  mechanical  analogy  we
would  like to show that there exist another solution of the problem
in question.

\section{ Flexural vibrations of the Timoshenko beam}
\setcounter{equation}0
To elucidate the analogy between the rigid string and the mechanical
vibrating   systems   we  consider  in  this  section  the  flexural
vibrations of the so-called Timoshenko beam.

In principle,  the flexural vibrations of three dimensional extended
objects such as rods or beams are described by the general equations
of  the  three  dimensional theory of elasticity [35].  However,  in
virtue of their complication this description is  not  suitable  for
practical   use.   Therefore,   one   has   to   employ   here  some
approximations.

If a rod or a beam is considered as an infinitely thin one (that is,
if we fully neglect its transverse sizes), then we obtain the string
described by the equation for the lateral deflection $y(x,t)$:
\begin{equation}
Ty''\,-\,\mu \ddot{y}\,=\,0.
\end{equation}
Here $T$  is  the  string tension and $\mu$ is the linear density of
the  string  matter.  As  it  was  to  be  expected,  none  of   the
characteristics of the transverse string sizes enter into (3.1).  By
taking into account the beam thickness effectively,  equation  (3.1)
is modified as~[24]
\begin{equation}
EIy''''\,-\,Ty'' \,+\,F\rho\ddot{y}\,=\,0,
\end{equation}
where $E$ is the Young's modulus,  $I$ is the momentum of inertia of
a cross section around the principal axis normal  to  the  plane  of
motion,  $F$  is  the  cross  section  area  and  $\rho$ is the mass
density.  In applications the case of the  absence  of  longitudinal
strength  ($T\,=\,0$) is frequently considered.  If it is really the
case,  then equation (3.2) is transformed into the  Bernoulli--Euler
equation
\begin{equation}
EIy''''\,-\,F\rho\ddot{y}\,=\,0.
\end{equation}

The effect  of  trasverse sizes of the beam leads to appearance,  in
equations (3.2) and (3.3),  of the higher  derivatives  as  compared
with  the  string  case (3.1).  The corresponding Lagrange densities
contain the $(y'')^2$  term,  but  the  problem  with  the  positive
definiteness  of the energy does not arise there.  Only the theories
with higher derivatives in time suffer from the above  problem.  The
model  of  flexural  vibrations of beams proposed at the begining of
our century by Timoshenko [24, 34] belongs to such theories. Besides
of  bending of the beam under the flexural vibrations the Timoshenko
model  takes  into   account   the   shear   deformations   of   its
elements.\footnote{Apart  this,  the inertia of gyration of the beam
cross sections is taken into account in the  Timoshenko  model  (the
Rayleigh correction [35]).  However,  this fact itself does not lead
to appearance of higher derivatives in  time  in  the  theory.}  Two
degrees  of  freedom  are  associated with each cross section of the
beam,  the deflection due to bending and that  due  to  shear.  This
duplication  of  the  number of degrees of freedom in the Timoshenko
model leads to the equation of the fourth order in time
\begin{equation}
EIy''''\,+\,F\rho\ddot{y}\,-\,\rho I\left(1\,+
\,\frac{E}{kG}\right)\ddot{y}''\,
+\,\rho I\frac{\rho}{kG}\stackrel{....}{y}\,=\,0.
\end{equation}
Here $G$  is the shear modulus and $k$ is the shear coefficient (the
phenomenological parameter depending on the  geometry  of  the  beam
cross section).

Equation (3.4) should be supplemented with the  boundary  conditions
at  the  ends $x_1\,=\,0,\,x_2\,=\,l$ of the beam.  In the following
for the sake of simplicity we shall consider the hinged-hinged beam,
where  both the flexure of the beam and its bending moment are equal
to zero
\begin{equation}
y(t,0)\,=\,y''(t,0)\,=\,0,\quad y(t,l)\,=\,y''(t,l)\,=\,0.
\end{equation}

The general  solution  of equation (3.4) and the boundary conditions
(3.5) has the form
\begin{equation}
y(t,x)\,=\,\sum_{n\neq 0}^{\infty}\sin\lambda_{n}x\left[q_{n1}(t)\,+
\,q_{n2}(t)
\right],
\end{equation}
where $\lambda_n\,=\,n\pi/l$, the functions $q_{ns}(t)\,=\,A_{ns}
\cos(\omega_{ns}t\,+\,\epsilon_{ns}),\,s\,=\,1,2$ are the normal
coordinates corresponding to  two  series  of  the  eigenfrequencies
$\omega_{ns}\,=\,\lambda_n   \sqrt{E/\rho}\omega_{*ns},\,s\,=\,1,2$,
respectively.  The  dimensionless  frequencies  $\omega_{*ns}$   are
defined by the formula
\begin{equation}
\left .
\begin{array}{c}
\omega_{*n1}^{2}\\
\omega_{*n2}^{2}
\end{array}
\right \}
\,=\,\frac{1}{2}\left [1\,+\,\xi\,+\,\frac {\xi}
{\lambda_{n}^{2}r^2}\,\mp\,
\sqrt{\left(1\,+\,\xi\,+\,\frac{\xi}
{\lambda_{n}^{2}r^2}\right)^2\,-\,4\xi\,}
\right ] ,
\end{equation}
where $\xi\,=\,kG/E$ is the  dimensionless  parameter,  $r$  is  the
radius  of  gyration of the beam cross section around principal axis
normal to the plane of motion, $r^2\,=\,I/F$.

When the  shear  modulus  $G$  tends  formally to the infinity,  the
Timoshenko equation (3.4) is reduced  to  the  Bernoulli--Euler  one
with the Rayleigh correction
\begin{equation}
EIy''''\,+\,\rho F\ddot{y}\,-\,\rho I\ddot{y}''\,=\,0.
\end{equation}
In this case the frequencies  of  the  first  series  (3.7)  in  the
Timoshenko theory tend to finite values
\begin{equation}
\omega_{*n1}^2\,\to \,\frac{\lambda_{n}^{2}r^2}{1\,+
\,\lambda_{n}^{2}r^2}
\end{equation}
and those of the second mode of oscillation go to infinity.

The Timoshenko   equation   (3.4)  and  the  corresponding  boundary
conditions (3.5)  can  be  derived  by  the  varying  the  following
Lagrangian density [36]
\begin{equation}
{\cal L}\,=\,\frac{1}{2}\left(\dot{y}^2\,-\,a_{1}y''^2\,-
\,a_{3}\ddot{y}^2\,+\,
a_{2}\ddot{y}y''\right).
\end{equation}
Here $a_i,\,i\,=\,1,2,3$ are the coefficients of  equation (3.4)
\begin{equation}
a_1\,=\,\frac{EI}{\rho F},\quad a_2\,=\,\frac{I}{F}\left(1\,+
\,\frac{E}{kG}\right),\quad
a_3\,=\,\frac{\rho I}{FkG}.
\end{equation}
Further, using  (2.13) one can define the canonical
variables
\[
q_1\,=\,y,\quad q_2\,=\,\dot{y},
\]
\[
p_1\,=\,\frac{\partial {\cal L}}{\partial \dot{y}}\,-
\,\frac{\partial}{\partial t}
\left(\frac{\partial {\cal L}}{\partial \ddot{y}}\right)\,=
\,\dot{y}\,+\,a_3
\ddot{y}\,-\,\frac{a_2}{2}\dot{y}',
\]
\begin{equation}
p_2\,=\,\frac{\partial {\cal L}}{\partial \ddot{y}}\,=
\,-a_3\ddot{y}\,+\,
\frac{a_2}{2}y''
\end{equation}
and construct the Ostrogradsky canonical Hamiltonian
\[
H\,=\,\frac{1}{2}\int\limits_{0}^{l}dx\left[2p_{1}q_2\,-
\,\frac{p_2^2}{a_3}\,-\,
q_2^2\,+\,\left(a_1\,-\,\frac{a_2^2}{4a_3}\right)q''^2_1\,+
\,\frac{a_2}{a_3}
p_{2}q''_1\right]\,=\,
\]
\begin{equation}
=\,\frac{1}{2}\int\limits_{0}^{l}dx\left(\dot{y}^2\,+\,
2a_3\dot{y}\stackrel{\ldots}{y}\,-\,a_2\dot{y}\dot{y}''\,-
\,a_3\ddot{y}^2\,+\,
a_{1} y''^{2} \right).
\end{equation}
This Hamiltonian  is  conserved  in  time  and it generates the time
translations $t \to t\,+\,  \Delta t$.  The  value  of  $H$  on  the
general  solution  (3.6)  is  the  energy  of  the  Timoshenko  beam
calculated according to Ostrogradsky
\begin{equation}
E_O\,=\,\frac{l}{4}a_3\sum_{n=1}^{\infty}\left(\omega_{n2}^2\,-
\,\omega_{n1}^2
\right)\left(\omega_{n1}^{2}A_{n1}^2\,-
\,\omega_{n2}^{2}A_{n2}^2\right).
\end{equation}
Thus, the  flexural vibrations with the amplitudes $A_{n2}$ give the
negative contribution to $E_{O}$ [36] because for all $n$'s we  have
from (3.7)
$$
\omega _{n2}^2\,-\,\omega _{n1}^2\,>\,0.
$$
Formula (3.14)  is  completely  equivalent  to  that  (2.16) for the
energy of the relativistic string  with  rigidity  in  the  harmonic
approximation.  In  spite  of  the  principal  difference  of  these
objects,  they suffer from the same lack of the lower energy  bound.
However,  in  the  case  of  the  flexural vibrations of beams there
exists the well definite notion of the mechanical  energy  which  is
always a positive quantity, of course.

The mechanical energy of a rod or a beam is a sum of the kinetic and
potential  ones  of  their  elements.  Let $y_1(t,\,x)$ be a lateral
deflection of the beam due to bending only and $y_2(t,\,x)$ be  that
due  to  shear.  In the Timoshenko model the kinetic energy contains
the contribution from the transverse motion of beam elements
\begin{equation}
T_{tr}\,=\,\frac{\rho FI}{2}\int\limits_{0}^{l}dx\,\dot{y}^2
\end{equation}
and that from the gyration of the beam cross section
\begin{equation}
T_{gyr}\,=\,\frac{\rho I}{2}\int\limits_{0}^{l}dx\,y'^2_1.
\end{equation}
Here $y(t,x)\,=\,y_1(t,x)\,+\,y_2(t,x)$   is   the   total   lateral
deflection of the beam.

According to  the Hooke law one can easily find the potential energy
of the flexural vibrations of the beam.  This energy consists of the
elastic energy of the bending deformations
\begin{equation}
V_{ben}\,=\,\frac{EI}{2}\int\limits_{0}^{l}dx\, y''^2_1,
\end{equation}
and that of the shear deformations
\begin{equation}
V_{sh}\,=\,\frac{kFG}{2}\int\limits_{0}^{l}dx\, y'^2_2.
\end{equation}

Joining together  formulae  (3.15)--(3.18)  we  obtain  the   action
functional of the Timoshenko model
\begin{equation}
W\,=\,\frac{\rho F}{2}\int\limits_{0}^{l}dx\,\left(\dot{y}^2\,+
\,r^2y'^2_1
\right)\,-\,\frac{EI}{2}\int\limits_{0}^{l}dx\,y''^2_1\,-
\,\frac{kFG}{2}
\int\limits_{0}^{l}dx\,y'^2_2.
\end{equation}
Variation of the action (3.19) gives  the  following  equations  for
$y_1(t,x)$ and $y_2(t,x)$
\begin{equation}
\frac{\rho}{E}\ddot{y}_1\,-\,y''_1\,=\,\frac{kG}{r^{2}E}y_2,
\end{equation}
\begin{equation}
\ddot{y}_2\,-\,\frac{kG}{\rho}y''_2\,=\,-\ddot{y}_1
\end{equation}
and the boundary conditions which take for the hinged--hinged beam
the form
\[
y(t,0)\,=\,y(t,l)\,=\,0,\quad y''(t,0)\,=\,y''(t,l)\,=\,0,
\]
\begin{equation}
\psi'(t,0)\,=\,\psi'(t,l)\,=\,0.
\end{equation}
Combining equations (3.20) and (3.21) one may obtain the  Timoshenko
equation (3.4) for the total lateral deflection $y\,=\,y_1\,+\,y_2$.

The sum of (3.15)--(3.18) is the total mechanical energy of flexural
vibrations of the Timoshenko beam
\begin{equation}
E\,=\,\frac{\rho F}{2}\int\limits_{0}^{l}dx\,\left(\dot{y}^2\,+
\,r^2\dot{\psi}^2
\right)\,+\,\frac{EI}{2}\int\limits_{0}^{l}dx\,\psi'^2\,+
\,\frac{kFG}{2}
\int\limits_{0}^{l}dx\,\left(y'\,-\,\psi\right)^2.
\end{equation}
Here $\psi(t,x)\,\equiv\,y'_1(t,x)$.   In   the    case    of    the
hinged--hinged   beam   we  have  the  general  solution  (3.6)  for
$y(t,\,x)$ and the analogous expansion for $\psi (t,\,x) $
\begin{equation}
\psi(t,x)\,=\,\sum_{n=1}^{\infty}\cos\lambda_{n}x
\left[\frac{k_{n1}}{l}q_{n1}
(t)\,+\,\frac{k_{n2}}{l}q_{n2}(t)\right]\,{,}
\end{equation}
where $k_{ns}/l$ are the amplitude ratios in  the  expansions  (3.6)
and (3.24) and
\begin{equation}
k_{ns}\,=\,n\pi\left(1\,-\,\xi^{-1}\omega_{*ns}^2\right),\quad s\,=
\,1,2,\quad
n\,=\,1,2,\ldots.
\end{equation}
Substituting (3.6)  and  (3.24) into (3.23) we obtain the expression
for   the   mechanical   energy   in   terms   of   the   amplitudes
$A_{ns},\,s\,=\,1,2$
\begin{equation}
E_{M}\,=\,\frac{l}{4}\sum_{n=1}^{\infty}\left[\left(1\,+
\,\frac{r^{2}k_{n1}^2}
{l^2}\right)\omega_{n1}^{2}A_{n1}^2\,+\,\left(1\,+
\,\frac{r^{2}k_{n2}^2}{l^2}
\right)\omega_{n2}^{2}A_{n2}^2\right]
\end{equation}
As it was to be expected,  the energy (3.26) is positive definite in
sign because of the positive definiteness of the original functional
(3.23).

So, in the Timoshenko  model  there  exists  the  mechanical  energy
positive   definite  in  sign  (formulae  (3.23),  (3.26))  and  the
Ostrogradsky energy unbounded from below (formulae (3.13),  (3.14)).
Both  these quantities are integrals of motion and they are mutually
related
\begin{equation}
E_{M}\,=\,E_{O}\,+\,\frac{l}{4}\left(\frac{a_{3}}{r}\right)^2
\sum_{n=1}^{\infty}
\frac{\left(\omega_{n2}^2\,-
\,\omega_{n1}^2\right)}{\lambda_n^2}\left[
\omega_{n2}^{4}A_{n2}^2\,-\,\omega_{n1}^{4}A_{n1}^2\right].
\end{equation}
However, the   mechanical   energy   (3.23)   in   contrast  to  the
Ostrogradsky energy (3.14) has quite a clear physical meaning.

\section{"Mechanical energy" of the  rigid string}
\setcounter{equation}0

The description of the rigid string  dynamics  (eqs.  (2.5),  (2.7),
(2.8)  and  (2.9))  is  in  many  respects  analogous to that of the
flexural vibrations of the  Timoshenko  beam  (eqs.  (3.4),  (3.20),
(3.21) and (3.22)). Indeed, both the objects can be described either
by one equation of the fourth  order  (equations  (2.5)  and  (3.4),
respectively)  or  by  two  equations of the second order (equations
(2.8), (2.9) and (3.20), (3.21) for the "partial" deflections). "The
material"  of  the  gluon tube in comparison with that of a beam has
very distinct mechanical properties,  of course. Therefore, in these
models  there  is  no  complete  identity  between the corresponding
equations.  But it is important that starting from  eqs.  (2.8)  and
(2.9)  in  the  rigid string model one may identify according to the
usual rules the energy corresponding to the mechanical  one  in  the
Timoshenko model.

For  equations (2.8) and (2.9) we have the standard Lagrangian
densities
\begin{equation}
{\cal L}_1\,=\,\frac{1}{2}\left(\dot{{\bf x}}_1^2\,-
\,{\bf x}'^2_1\right),\,
{\cal L}_2\,=\,\frac{\epsilon}{2}\left(\dot{{\bf x}}^2_2\,-
\,{\bf x}'^2_2\right)
\,-\,\frac{{\bf x}^2_2}{2}.
\end{equation}
The total energy is defined by the formula
\begin{equation}
E_{M}\,=\,\frac{1}{2}\int\limits_{0}^{\pi}d\sigma\,
\left(\dot{{\bf x}}^2_1\,+\,
{\bf x}'^2_1\right)\,+\,\frac{1}{2}\int\limits_{0}^{\pi}d\sigma\,
\left(
\dot{{\bf x}}^2_2\,+\,{\bf x}'^2_2\,+\,{\bf x}^2_2\right).
\end{equation}
Substituting the general solution (2.10) into (4.2) one finds
\begin{equation}
E_{M}\,=\,\frac{{\bf P}^2}{2M}\,+
\,\frac{a\hbar}{2}\sum_{n=1}^{\infty}\left(
{\ab }_n^{*}{\ab }_n\,+\,{\ab }_{n}{\ab }^{*}_n\right)\,+\,
\frac{a\hbar}{2}\sum_{n=1}^{\infty}\left({\bb }^{*}_{n}{\bb }_n\,+\,
{\bb }_{n}{\bb }^{*}_n\right).
\end{equation}

As it  was to be expected,  the mechanical energy (4.3) in the rigid
string model is the quatity positive definite  in  sign.  Obviuosly,
this property of the energy also holds at the quantum level.  Taking
account of zero--point oscillations one  may  write  the  mechanical
energy of the rigid string as follows

\[
E_{M}\,=\,\frac{{\bf P}^2}{2M}\,+
\,a\hbar\sum_{n=1}^{\infty}\omega_{n1}\left(
{\bf a}^{+}_{n}{\bf a}_n\,+\,\frac{D-2}{2}\right)\,+
\]
\begin{equation}
+\,a\hbar\sum_{n=1}^{\infty}
\omega_{n2}\left({\bf b}^{+}_{n}{\bf b}_n\,+\,\frac{D-2}{2}\right).
\end{equation}
In this  case  all  string  states in the Fock space are positive in
norm,  hence the above mentioned problem with violation of unitarity
does not arise here.

\section{Conclusion}

Thus in the framework of the rigid string model we have  shown  that
one can construct, for this object, a positive definite "mechanical"
energy instead of the  Ostrogradsky  energy  unbounded  from  below.
Obviously,  the  same  can  be  done  for any field model describing
extended objects at the classical level.  An appealing future of our
approach  is  the absence of any constraints on the physical degrees
of freedom introduced  "by  hand"  in  some  other  papers  on  this
subject.  This  enables  one  to construct a complete quantum theory
instead of the truncated one.  Further,  at the  quantum  level  the
problems  with negative norm states and the loss of unitarity do not
arise.

On the other hand,  the energy constructed according to Ostrogradsky
generates  the  time translations,  but the mechanical one does not.
Therefore,  a sole difficulty which can occur here is to  prove  the
relativistic invariance of such theories by making use of the notion
of the mechanical energy.
\vskip1cm
{\large \bf Acknowledgments}
This work is supported by the  Russian  Foundation  for  Fundamental
Research through project 93-02-3972.  One of the authors, V.\ V.\ N.
would like to thank N.\ R.\ Shvetz for the valuable  discussions  at
the early stage of this work.

\vfill
\begin{center}
Received by Publishing Department \\
on May 28, 1993
\end{center}
\end{document}